\documentclass{raa}            

\usepackage{graphicx,times}             
\usepackage{natbib}
\usepackage{amssymb,amsmath}
\usepackage{gensymb}
\bibpunct{(}{)}{;}{a}{}{,}

\usepackage[a4paper=true,dvipdfm=true,pagebackref=true]{hyperref}
\hypersetup{colorlinks = true, linkcolor = green, anchorcolor = red, citecolor = blue, filecolor = red, pagecolor = red, urlcolor = red}

\begin{document}

\title{Near-resonance Tidal Evolution of the Earth-Moon System Influenced by Orbital-scale Climate Change}

\volnopage{Vol.0 (20xx) No.0, 000--000}      
\setcounter{page}{1}          

\author{Nan Wang
    \inst{}
\and Zhi-Guo He
    \inst{}}

\institute{Ocean College, Zhejiang University, Hangzhou 310058, China; {\it hezhiguo@zju.edu.cn}\\
\vs\no
   {\small Received~~20xx month day; accepted~~20xx~~month day}}

\abstract{
%
We build a conceptual coupled model of the climate and tidal evolution of the Earth-Moon system to find the influence of the former on the latter. An energy balance model is applied to calculate steady-state temperature field from the mean annual insolation as a function of varying astronomical parameters. A harmonic oscillator model is applied to integrate the lunar orbit and Earth's rotation with the tidal torque dependent on the dominant natural frequency of ocean. An ocean geometry acts as a bridge between temperature and oceanic frequency.
On assumptions of a fixed hemispherical continent and an equatorial circular lunar orbit, considering only the 41\,kyr periodicity of Earth's obliquity $\varepsilon$ and the $M_2$ tide, simulations are performed near tidal resonance for $10^6$\,yr.
It is verified that the climate can influence the tidal evolution via ocean. Compared with the tidal evolution with constant $\varepsilon$, that with varying $\varepsilon$ is slowed down; the Earth-Moon distance oscillates in phase with $\varepsilon$ before the resonance maximum but exactly out of phase after that; the displacement of the oscillation is in positive correlation with the difference between oceanic frequency and tidal frequency.
%
}

\titlerunning{Tidal Evolution Influenced by Climate Change}

\maketitle

\section{Introduction}\label{sec-intro}

The tidal evolution of the Earth-Moon system is a classic problem, but has not yet been fully solved.
The general trend of the tidal evolution has long been confirmed: since the Moon was born 4.5 Gyr ago (\citealt{Halliday-2008}), it has been receding from the Earth with tidal energy dissipated as heat, and Earth's rotation has been slowing down with angular momentum transferred to the lunar orbit (\citealt{Murray-1999}).
When one natural frequency (also called a normal mode) of the ocean and one tidal forcing frequency, both varying over geologic time, come close to each other, the ocean tide gets excited and the dissipation is largely enhanced. This phenomenon of "tidal resonance" speeds up tidal evolution. Currently, the $M_2$ (semidiurnal tide) resonance in the ocean contributes a dissipation of 2.4\,TW (\citealt{Munk-1997}) to the total of 3.7\,TW (\citealt{Munk-1998}), which is so abnormally high that extrapolating into the past unrealistically puts the Moon where it was born as recently as $\sim 2$\,Gyr ago (e.g, \citealt{Touma-1994,Bills-1999}). Because of the uncertainty that tidal resonance brings, to quantitatively reconstruct the lunar orbit, the history of oceanic natural frequencies has to be acquired.

The natural frequencies of the ocean are determined by its geometry (position, shape and depth), which is associated with climate change and continental drift.
In order to simulate tidal evolution driven by ocean tide, \cite{Hansen-1982} used Laplace's tidal equations to determine the oceanic tidal torque for two ocean/land geographies, but not only the geography but also the ocean depth remained unchanged in simulations. \cite{Webb-1980,Webb-1982-GJI-68,Webb-1982-GJI-70} developed a model of average ocean, a statistical average over many hemispherical oceans centered at various positions relative to Earth's axis, to take the change in ocean geometry due to continental drift into account, whereas the ocean shape and depth were constant. \cite{Kagan-1994} built a stochastic model, which considered the effect of continental drift as explicit fluctuations in oceanic natural frequencies.
By solving the timescale problem (e.g., \citealt{Goldreich-1966}), those ocean modelers demonstrated the importance of the ocean to tidal evolution of the Earth-Moon system, but their results are still qualitative without further improvement for decades, and none has involved climate.
The timescale of continental drift is $10^8$\,yr (\citealt{Murray-1999}), and that of orbital-scale climate change is $\sim 10^5$\,yr (\citealt{Berger-2012}). Considering only continental drift means the secular effect of climate influence can be neglected, which is based on insufficient evidence. Hence, we simulate the tidal evolution for $10^6$\,yr, so that the influence of climate can be investigated and continental drift can be reasonably neglected.

Natural quasi-periodicities of the climate over timescales in a vast range have been discovered through geological records. On the orbital-scale of $\sim 10^5$\,yr, the glacial-interglacial cycle dominates (\citealt{Berger-2012}). During a glacial, the ice sheets extend towards the equator and the sea level drops; whereas during an interglacial, the ice sheets shrink towards the poles and the sea level rises. According to the Milankovitch theory, that results from the secularly varying orbit and rotation of the Earth perturbed by the Sun and other planets, and the subsequent variation of insolation distribution on the terrestrial surface (e.g., \citealt{Berger-1988}). This effect on climate of the astronomical parameters (Earth's eccentricity, obliquity and climatic precession) is called astronomical forcing. The change in sea level, which the glacial-interglacial cycle is accompanied by, alters the natural frequencies of the ocean and then the state of tidal resonance. For instance, in the last glacial maximum $\sim$19--26\,kyr ago, the sea level drop of about $130$\,m (\citealt{Yokoyama-2000,Clark-2009,Lambeck-2014}) leads to considerably higher dissipation than at present (\citealt{Thomas-1999,Egbert-2004,Griffiths-2009,Green-2009}). Therefore, it is the ocean that acts as the bridge between climate and tidal evolution.

In this work, a coupled model of climate and tidal evolution is proposed. An energy balance model (\citealt{Sellers-1969,Budyko-1969}) is applied to simulate the climate in response to astronomical forcing. It is a kind of conceptual model focusing on major climate components and interactions. Though very simplified, it is capable of reproducing the glacial variability with ice sheet involved (\citealt{Huybers-2008,McGehee-2012}) and frequently used to study the climate stability (see review in \cite{North-1984} for early studies; \citealt{Lin-1990,Wagner-2015}). A harmonic oscillator model (\citealt{Munk-1968,Murray-1999}) is applied to integrate the lunar orbit and Earth's rotation with the oceanic natural frequency given. It is also a conceptual model, where the response of the ocean to the tidal forcing is compared to that of a harmonic oscillator, and is capable of providing a realistic timescale of tidal evolution (\citealt{Hansen-1982,Kagan-1994}). It suits the case when there is one dominant oceanic natural frequency and one dominant tidal forcing frequency. In addition, a simplified ocean geometry is assumed to obtain the natural frequency from temperature field.

As a preliminary effort to study the influence of climate on tidal evolution, we aim at verifying the existence of the influence and qualitatively observing its nature and mechanism. Therefore, our idealized model and simulation time ($10^6$\,yr) are appropriate. The period of interest is when the ocean and tidal forcing are near resonance (but not at the resonance maximum), so that the influence can be amplified. Section~\ref{sec-model} describes the coupled model and the numerical method, and Section~\ref{sec-results} exhibits the results of two sets of simulations in pre- and post-resonance times. A discussion about simulation results and potential improvements is in Section~\ref{sec-discussion}.

\section{Model and Method}\label{sec-model}

\subsection{Climate Model}\label{sec-climate}

\subsubsection{Steady-state temperature field}

The climate can be altered by ocean/land geography. To study climate change resulting from astronomical forcing, a simple geography is used and taken to be invariant.
It is assumed that a single spherical-cap continent is centered at the North Pole, extending to latitude $\varphi_{\rm{l}}$, and the rest of Earth's surface is covered by ocean. Such a geography is similar to what \cite{Mengel-1988} used.
Neglecting the vertical structure of the atmosphere for this zonally symmetric planet, a one-dimensional climate model can be applied.

Considering horizontal thermal diffusion, outgoing infrared radiation as well as the solar heating being the only external forcing, an energy balance model leads to this governing equation (\citealt{North-2017})
\begin{equation}\label{eq-EBM}
    C(\varphi) \frac{\partial T(\varphi,t)}{\partial t} - \frac{1}{\cos{\varphi}} \frac{\partial}{\partial \varphi} [\frac{D}{\cos{\varphi}} \frac{\partial T(\varphi,t)}{\partial \varphi}] - [A + B T(\varphi,t)] = W(\varphi) \tilde{\alpha}(\varphi).
\end{equation}
The climate at time $t$ is just characterized by the temperature field $T(\varphi)$ on the surface. In the first term on the left side, $C$ is the effective heat capacity and controls the climate response to perturbations (relaxation time $\tau = C/B$). The capacity over the ocean $C_{\rm{w}}$
is larger than that over the land $C_{\rm{l}}$,
and so is the relaxation time for ocean $\tau_{\rm{w}}$ (a few years) than that for land $\tau_{\rm{l}}$ (a month). According to the geography assumed,
\begin{equation}
	C(\varphi)=
	\begin{cases}
		C_{\rm{l}},	& (\varphi > \varphi_{\rm{l}})\\
		C_{\rm{w}}.	& (\varphi < \varphi_{\rm{l}})
	\end{cases}
\end{equation}
The second term involving the thermal diffusion coefficient $D$ allows the heat transport from warm areas to cool. The third term allows the infrared radiation to space and $A$ and $B$ are empirical coefficients from satellite observations. The term on the right side determines the solar radiation absorbed. The insolation function $W(\varphi)$ gives the latitudinal distribution of the solar radiation flux delivered to the surface. It is dependent on Earth's orbital status (Sect.~\ref{sec-insolation}). The coalbedo $\tilde{\alpha}(\varphi)$ gives the fraction of radiation absorbed by the surface. Its mean annual form is well represented by
\begin{equation}
	\tilde{\alpha}(\varphi)= \tilde{\alpha}_0 + \tilde{\alpha}_2 P_2(\sin\varphi),
\end{equation}
where constants $\tilde{\alpha}_0$ and $\tilde{\alpha}_2$ are from satellite observations, and the second-order Legendre polynomial $P_2(\sin\varphi) = (3\sin^2\varphi - 1)/2$.
The values of the constants mentioned above and the references that provide them are listed in Table~\ref{tab-values}.
%
Solving Equation~\ref{eq-EBM} also needs the boundary condition
\begin{equation}
	\frac{\partial T}{\partial \varphi}=0,	\quad (\varphi=\pm 90\dg)
\end{equation}
implying that there is no net heat flux into the poles.

On the assumption of energy balance (the energy absorbed is equal to the energy lost), for every given insolation function $W(\varphi)$, there exists a corresponding steady-state solution $T^{\rm{s}}(\varphi)$, which any temperature field $T(\varphi)$ of a different profile relaxes to after a time comparable to $\tau$. The relaxation time $\tau$ of the climate system is much smaller than the astronomically driven period of mean annual $W(\varphi)$. Therefore, $W(\varphi)$ can be taken as invariant while $T(\varphi)$ is evolving towards the steady state $T^{\rm{s}}(\varphi)$.

For a steady-state temperature field $T^{\rm{s}}(\varphi)$, the iceline $\varphi_{\rm{f}}$, the edge of the permanent ice cap, is determined by
\begin{equation}\label{eq-T}
    T^{\rm{s}}(\varphi_{\rm{f}}) = T_{\rm{f}},
\end{equation}
where the mean annual isotherm $T_{\rm{f}} = -10\celsius$ (\citealt{North-1979}). We note that although the iceline is defined, expressions of the capacity $C(\varphi)$ and coalbedo $\bar{\alpha}(\varphi)$ are not influenced by it, that is, the ice-albedo feedback is not included in the present work. The iceline position only influences the sea level (Sect.~\ref{sec-ocean}).

\subsubsection{Insolation distribution}\label{sec-insolation}

Given the timescale of interest, the mean annual version of the insolation function $W(\varphi)$ is used, with seasonal variation averaged. It is dependent on Earth's semimajor axis $a$, eccentricity $e$ and obliquity $\varepsilon$ (\citealt{Loutre-2004}). Assuming the Sun is a point source, \cite{Berger-2010} deduced with elliptical integrals the total energy available during any time interval of one year on a given unit surface. Based on their result, we deduce the expression of the mean annual insolation function as
\begin{equation}\label{eq-W}
    W(\varphi, a, e, \varepsilon)=
    \begin{cases}
        \frac{L_{\odot} \cos\varphi}{2 \pi^3 a^2 \sqrt{1-e^2}} [E(\frac{\sin\varepsilon}{\cos\varphi}) + \tan^2\varphi K(\frac{\sin\varepsilon}{\cos\varphi}) - \tan^2\varphi \cos^2\varepsilon \Pi(\sin^2\varepsilon, \frac{\sin\varepsilon}{\cos\varphi})],  & (|\varphi| \in [0\dg, 90\dg - \varepsilon)) \\
        \frac{L_{\odot} \sin\varepsilon}{2 \pi^3 a^2 \sqrt{1-e^2}} [E(\frac{\cos\varphi}{\sin\varepsilon}) + \cot^2\varepsilon K(\frac{\cos\varphi}{\sin\varepsilon}) - \sin^2\varphi \cot^2\varepsilon \Pi(\cos^2\varphi, \frac{\cos\varphi}{\sin\varepsilon})],  & (|\varphi| \in (90\dg - \varepsilon, 90\dg)) \\
        \frac{L_{\odot}}{2 \pi^3 a^2 \sqrt{1-e^2}} [\sin\varepsilon + \frac{\cos^2\varepsilon}{2} \ln(\frac{1 + \sin\varepsilon}{1 - \sin\varepsilon})],   & (|\varphi| = 90\dg - \varepsilon) \\
        \frac{L_{\odot}}{4 \pi^2 a^2 \sqrt{1-e^2}} \sin\varepsilon.  & (|\varphi| = 90\dg)
    \end{cases}
\end{equation}
The solar luminosity $L_{\odot}$ is a constant, and $K$, $E$ and $\Pi$ are the first, second and third complete elliptical integrals, respectively.
This expression is valid for $0\dg < \varepsilon < 90\dg$ and $0 < e < 1$.

Variation of $W(\varphi)$ results from those of the astronomical parameters $e$ and $\varepsilon$ (mainly over a timescale of $\sim 10^5$\,yr). Based on the numerical solution for Earth's orbit (e.g, \citealt{Laskar-1988}), $e$ and $\varepsilon$ can be expressed in trigonometric form as quasi-periodic functions of $t$: ${\textrm{approximation}} + \sum{ \{ {(\textrm{amplitude}})_i \cos{[{(\textrm{frequency}})_i t + {(\textrm{phase})}_i]} \} }$ (\citealt{Berger-2012}).
In this work, we only consider the most important term of $\varepsilon$, whose period is 41\,kyr, in order to avoid the compound influence of its multiple terms and the complicated effect of $e$ and $\varepsilon$ simultaneously varying. Thus, $e = \bar{e}$ and
\begin{equation}\label{eq-orb}
    \varepsilon (t) = \bar{\varepsilon} + \Delta \varepsilon \cos(\gamma t + \psi),
\end{equation}
where values of the approximation $\bar{\varepsilon}$, the amplitude $\Delta \varepsilon$ and the frequency $\gamma$ given by \cite{Berger-1991} are used, which are valid over 1--3\,Myr (\citealt{Berger-1992}). However, the phase $\psi$ acts as a controllable parameter in our simulations to manifest the influence of itself.

\subsection{Ocean Geometry Model}\label{sec-ocean}

An ocean geometry model is needed to connect climate and tidal evolution. An ocean function with a value of 1 over ocean and 0 over land can be defined to characterize the geography modeled in Section~\ref{sec-climate}
\begin{equation}\label{eq-h}
    h(\varphi) =
    \begin{cases}
        1,  &  (\varphi \leq \varphi_{\rm{l}})  \\
        0.  &  (\varphi > \varphi_{\rm{l}})
    \end{cases}
\end{equation}

The iceline $\varphi_{\rm{f}}$ divides the water on Earth's surface into two reservoirs, i.e., sea water in the ocean basin and ice sheet on the part of continent north of the iceline. It is further assumed that the depth of the ocean basin $h_{\rm{b}}$, the depth of the sea water $h_{\rm{sw}}$ and the thickness of the ice sheet $h_{\rm{is}}$ are all uniform. Sea ice is not considered.  
By simply taking the volume of sea water as its depth times the area of the ocean basin and the volume of the ice sheet as its thickness times the area of ice cover, conservation of mass gives rise to
\begin{equation}\label{eq-h_sw}
    h_{\rm{sw}} = h_{\rm{b}} - \frac{\rho_{\rm{is}} h_{\rm{is}} (1 - \sin \varphi_{\rm{f}})}{\rho_{\rm{sw}} (1 + \sin \varphi_{\rm{l}})},
\end{equation}
where $\rho_{\rm{is}}$ and $\rho_{\rm{sw}}$ are densities of ice sheet and sea water, respectively, and $h_{\rm{b}}$ is equal to the maximum of $h_{\rm{sw}}$ which corresponds to the condition $\varphi_{\rm{f}} = 90\dg$.

The oceanic natural frequency $\sigma$ is determined by the geometry of the ocean. On the assumption of half-wavelength resonance, the ocean basin is simplified as a closed square and $\sigma$ can be estimated to be
\begin{equation}\label{eq-omega_0}
    \sigma = \frac{\pi \sqrt{g h_{\rm{sw}}}}{l},
\end{equation}
where $l$ is the ocean width and $g$ is the gravitational acceleration. Typical values of $h_{\rm{is}}$, $\bar{h}_{\rm{sw}}$ (mean of $h_{\rm{sw}}$) and $l$ are set (Table~\ref{tab-values}).
The advantage in setting $l$ to a typical value instead of deriving it from the modeled geography is that a realistic $\sigma$ value can be obtained.
Because the geography is invariant, the oceanic frequency is only dependent on climate, i.e., $\sigma$ only varies with $h_{\rm{sw}}$. On the other hand, $\sigma$ is yet needed in the calculation of tidal torque coefficient $Z$ (Sect.~\ref{sec-torque}). The climate and tidal evolution are thus connected.

\subsection{Tidal Evolution Model}\label{sec-tidal}

\subsubsection{Orbital parameters}\label{sec-orb}

A two-body system consisting of the Earth and Moon with a circular orbit in the Earth's equatorial plane is considered.
With the Moon's rotational angular momentum neglected, conservation of angular momentum leads to
\begin{equation}\label{eq-H}
    I \Omega + M_{\rm{r}} n r^2 = H.
\end{equation}
The first term on the left side is Earth's rotational angular momentum, where $I$ is Earth's rotational inertia and $\Omega$ is Earth's rotational speed. The second term is the lunar orbital angular momentum, where $n$ is the lunar angular orbital speed, $r$ is the Earth-Moon distance, and the reduced mass $M_{\rm{r}} = M_{\oplus} M_{\rm{M}}/(M_{\oplus} + M_{\rm{M}})$ ($M_{\oplus}$ and $M_{\rm{M}}$ are masses of Earth and Moon, respectively). The total angular momentum $H$ is constant.
There are three orbital parameters, $r$, $n$ and $\Omega$, characterizing the state of tidal evolution. Besides Equation~\ref{eq-H}, $n$ and $r$ are also linked by Kepler's third law,
\begin{equation}\label{eq-nr}
    n^2 r^3 = G(M_{\oplus} + M_{\rm{M}}),
\end{equation}
where $G$ is the gravitational constant. Therefore, knowledge of one of $r$, $n$ and $\Omega$ is equivalent to that of them all.

The evolution of $\Omega$ is determined by tidal torque $L$ (Sect.~\ref{sec-torque}), which arises because the Earth, carrying its tidal bulge, rotates faster than the Moon orbiting it ($\Omega > n$)
\begin{equation}\label{eq-Omega}
    I \frac{{\rm{d}} \Omega}{{\rm{d}} t} = - L.
\end{equation}
The tidal torque acts to decrease $\Omega$, resulting in a transfer of angular momentum from Earth's rotation to lunar orbital motion and a dissipation of energy in Earth.

\subsubsection{Tidal torque}\label{sec-torque}

A tide is raised on an elastic body when this body is distorted in the gravity field of another. In our model, the solid Earth and Moon are taken as rigid spheres, while the ocean is a thin deformable layer partially covering the solid Earth. Therefore, only the tidal torque exerted on the Moon by the distorted ocean is present. Furthermore, only the semidiurnal tide $M_2$, the dominant tidal constituent at present, is considered in this work for simplification.

\cite{Hansen-1982} derived an approximate expression of the secular variation of the tidal constituent torque after averaging on a short timescale (monthly and yearly) and neglecting terms higher than $(R_{\oplus}/r)^8$. For the assumed equatorial lunar orbit, that expression becomes
\begin{equation}\label{eq-L}
    L(\sigma, \omega, r) = - L_* \cdot (r_{\rm{p}}/r)^6 \cdot {\rm{Im}} (Z(\sigma, \omega)),
\end{equation}
where the constant $L_*$, which carries the dimensionality of the torque, is
\begin{equation}
	L_* = \frac{6}{5} \pi \rho_{\rm{sw}} R_{\oplus}^2 \cdot G M_{\oplus} \cdot (M_{\rm{M}}/M_{\oplus})^2 \cdot (R_{\oplus}/r_{\rm{p}})^6,
\end{equation}
$r_{\rm{p}}$ is the present Earth-Moon distance and $R_{\oplus}$ is the radius of Earth.

Additionally, the torque coefficient $Z$ in Equation~\ref{eq-L} is a second-degree spherical harmonic expansion coefficient of the complex tidal elevation response function. For a static ocean tide whose shape is the same as that of the tide potential, $Z$ degenerates into a static torque coefficient
\begin{equation}\label{eq-Z_static}
    Z^{\rm{static}} = [\langle |Y|^2 h \rangle - |\langle Y h \rangle|^2/\langle h \rangle]/\langle |Y|^2 \rangle,
\end{equation}
where the complex spherical harmonic $Y = P_2^{2}(\cos\theta) {\rm{e}}^{{\rm{i}} 2 \lambda}$ for $M_2$ tide (the unnormalized associated Legendre function $P_2^{2}(\cos\theta) = 3 \sin^2\theta$, $\theta$ is colatitude and $\lambda$ is longitude), the ocean function $h$ is defined in Equation~\ref{eq-h}, and the angled brackets imply an areal integration over the global surface.

For a dynamic ocean tide, following \cite{Hansen-1982}, the harmonic oscillator model is adopted to derive $Z$. A driven harmonic oscillator can be described as
\begin{equation}
    \frac{{\rm{d}}^2 \zeta}{{\rm{d}} t^2} + \delta \sigma \frac{{\rm{d}} \zeta}{{\rm{d}} t} + \sigma^2 \zeta = \sigma^2 \zeta_{\rm{m}}^* {\rm{e}}^{{\rm{i}} \omega t},
\end{equation}
where $\zeta$ is the displacement from equilibrium, $\delta$ is the frictional resistance coefficient, $\sigma$ is the natural frequency of the oscillator, $\omega$ is the frequency of the external force and $\zeta_{\rm{m}}^*$ is the limit of maximal displacement as $\omega$ approaches 0. Its steady-state solution is $\zeta = \zeta_{\rm{m}} {\rm{e}}^{{\rm{i}} \omega t}$, where the displacement amplitude is
\begin{equation}
    \zeta_{\rm{m}} = \frac{\zeta_{\rm{m}}^*}{1- (\omega/\sigma)^2 + {\rm{i}} \delta (\omega/\sigma)}.
\end{equation}
In an analogy with ocean tides, $\zeta$ is taken as the tidal elevation in the ocean, $\sigma$ is the oceanic natural frequency, $\omega$ is the tidal forcing frequency, $\zeta_{\rm{m}}^* = Z^{\rm{static}}$ and $\zeta_{\rm{m}} = Z$. The tidal torque coefficient is thus derived
\begin{equation}\label{eq-Z}
    Z(\sigma, \omega) = \frac{Z^{\rm{static}}}{1- (\omega/\sigma)^2 + {\rm{i}} \delta (\omega/\sigma)}.
\end{equation}

This tidal torque coefficient $Z$ varies with $\sigma$ and $\omega$, for $Z^{\rm{static}}$ absolutely determined by geography is constant in our model. The oceanic natural frequency $\sigma$ is dependent on the state of ocean (Sect.~\ref{sec-ocean}), while the tidal forcing frequency for $M_2$ tide is
\begin{equation}\label{eq-omega}
    \omega(\Omega, n) = 2(\Omega - n).
\end{equation}
If $\sigma$ and $\omega$ are close enough (not strictly equal for a nonzero $\delta$), ${\rm{Im}} (Z)$ will be largely enhanced and so will $L$. Tidal evolution then speeds up, and that is when a tidal resonance is considered to occur.

\subsection{Method of Solution}\label{sec-method}

\begin{table}
\begin{center}
\caption[]{Values of Model Constants.}\label{tab-values}
\begin{tabular}{lll}
    \hline\noalign{\smallskip}
    Constant                    & Value                     & Reference        \\
    \hline\noalign{\smallskip}
    $A$                         & 218\,W~m$^{-2}$           & \cite{North-2017} \\
    $B$                         & 1.90\,W~m$^{-2}$~K$^{-1}$ & \cite{North-2017} \\
    $D$                         & 0.67\,W~m$^{-2}$~K$^{-1}$ & \cite{North-2017} \\
    $C_{\rm{l}}$                & 0.08\,yr$ \cdot B$        & \cite{Lin-1990}   \\
    $C_{\rm{w}}$                & 4.80\,yr$ \cdot B$        & \cite{Lin-1990}   \\
    $\tilde{\alpha}_0$          & 0.68                      & \cite{North-2017} \\
    $\tilde{\alpha}_2$          & $-0.20$                   & \cite{North-2017} \\
    $\bar{e}$                   & 0.028707                  & \cite{Berger-1978-JAS}\\
    $\bar{\varepsilon}$         & 23.333410\dg              & \cite{Berger-1991}\\
    $\Delta \varepsilon$        & $-1969.00"$               & \cite{Berger-1991}\\
    $\gamma$                    & 31.54068"~yr$^{-1}$       & \cite{Berger-1991}\\
    $\varphi_{\rm{l}}$          & 0\dg                      & present work      \\
    $T_{\rm{f}}$                & $-10\celsius$             & \cite{North-1979} \\
    $\rho_{\rm{is}}$            & 0.917\,g~cm$^{-3}$       & \cite{Haynes-2017}\\
    $\rho_{\rm{sw}}$            & 1.037\,g~cm$^{-3}$        & \cite{Hansen-1982}\\
    $h_{\rm{is}}$               & 2\,km                     &                   \\
    $\bar{h}_{\rm{sw}}$         & 4\,km                     &                   \\
    $l$                         & 4000\,km                  &                   \\
    $\delta$                    & 0.092                     & present work      \\
    \noalign{\smallskip}\hline
\end{tabular}
\end{center}
\tablecomments{\textwidth}{
Values given with no reference are typical in reality.}
\end{table}

\begin{figure}
\centering
\includegraphics[height=10cm]{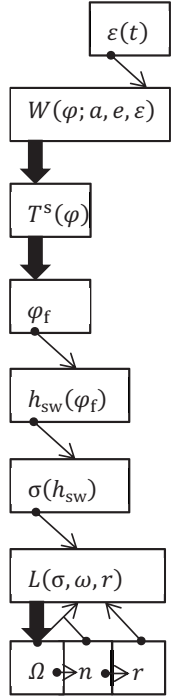}
\caption{Procedure for calculating model variables. The thick arrows represent numerical methods, while the thin arrows signify substitution into analytical expressions.}
\label{fig-procedure}
\end{figure}

The values of constants involved in our model are listed in Table~\ref{tab-values}. Because of the simplification of our model, whether their values are precise or not does not affect our qualitative conclusions.
Numerical simulations are executed for different initial Earth-Moon distances and phases of Earth's obliquity. The initialization of simulations will be presented in Section~\ref{sec-results}. The following is the procedure carried out for any instant after the initial time as shown in Figure~\ref{fig-procedure}.

In every calculation loop characterized by $t$, obliquity $\varepsilon$ is the first to be obtained. The term with the biggest amplitude 0.547\dg (\citealt{Berger-1991}) in its trigonometric expansion is used to get $\varepsilon$ (Eq.~\ref{eq-orb}). The corresponding period is 41090\,yr. Then, the insolation function $W(\varphi)$ is derived from $\varepsilon$ using Equation~\ref{eq-W}, with $a$ fixed at its present value and $e$ fixed at its approximation over the last few million years $\bar{e}$.

An numerical method is applied to derive the steady-state temperature field $T^{\rm{s}}(\varphi)$ from $W(\varphi)$.
Specifically, the differential equation of $T$ (Eq.~\ref{eq-EBM}) is discretized by centered finite difference method.
The latitude is discretized in intervals of $\Delta \varphi = \pi/180$, while the time step is given by $\Delta t/2$, where the propagation time $\Delta t = (\Delta \varphi^2/2) (C_{\rm{l}}/D)$. Stepping forward in time from the initial condition $T(\varphi) = 10\celsius$, the iteration does not cease until the relative error of temperature is less than $10^{-6}$, which happens after no more than $3.5 \tau_{\rm{w}} = 17$\,yr in our simulations. Thus, a steady-state solution is considered to be reached.
Because it is instantly reached compared to the time step in integration for tidal evolution ($\sim 10^3$\,yr), the steady-state solution $T^{\rm{s}}(\varphi)$ solved with $W(\varphi)$ given at time $t$ is just taken as the temperature field at that instant.
We note that the topography adopted is a hemispherical continent ($\varphi_{\rm{l}} = 0\dg$), and the nonlinearity in $C(\varphi)$ does not introduce any convergence problems in our procedure.

Defining $T_{\rm{f}} = -10 \celsius$, the iceline $\varphi_{\rm{f}}$ (Eq.~\ref{eq-T}) is then quickly located by linearly interpolating the adjoined latitudes where temperatures are found to just enclose $T_{\rm{f}}$.
The sea water depth $h_{\rm{sw}}$ is calculated from $\varphi_{\rm{f}}$ using Equation~\ref{eq-h_sw}, where the maximal depth $h_{\rm{b}}$ is set (in the first loop) to what ensures that the mean depth $\bar{h}_{\rm{sw}} = 4$\,km (equal to $h_{\rm{sw}}$ in the first loop, given $\psi = \pm 90\dg$) during simulations. The oceanic natural frequency $\sigma$ is calculated from $h_{\rm{sw}}$ using Equation~\ref{eq-omega_0}, where the ocean width is set to $l = 4000$\,km so that the mean natural frequency $\bar{\sigma} = 1.555 \times 10^{-4}$\,rad~s$^{-1}$ is near the present $M_2$ resonance.

After $\sigma$ is obtained, a Runge-Kutta-Fehlberg method is applied to integrate the differential equation of Earth's rotational speed $\Omega$ (Eq.~\ref{eq-Omega}). As pointed out in Section~\ref{sec-orb}, the lunar orbital parameters $\Omega$, $n$ and $r$ are related by Equation~\ref{eq-H} and \ref{eq-nr}, so $n$ and $r$ can be known with $\Omega$ given. In Equation~\ref{eq-H}, the total angular momentum $H = 3.442 \times 10^{34}$\,kg~m$^2$~s$^{-1}$ which is determined by substituting $\Omega$, $n$ and $r$ with present values. The tidal frequency $\omega$ is then also known because of its dependence on $\Omega$ and $n$ (Eq.~\ref{eq-omega}).
Therefore, within each time step of the Runge-Kutta-Fehlberg integrator, while $\sigma$ is held constant, $L$ can be obtained with $r$ and $\omega$ by using Equation~\ref{eq-Z_static}, \ref{eq-Z} and \ref{eq-L} in turn, following the calculation of $r$ and $\omega$ from the coinstantaneous $\Omega$. In these equations, the static torque coefficient $Z^{\rm{static}} = 0.5$ for a hemispherical continent, the frictional coefficient $\delta$ is set to 0.092 in order to ensure a realistic timescale of tidal evolution based on some test simulations and the dimensionality is calculated to be $L_* = 1.998 \times 10^{17}$\,N~m.
%

Given $\Omega$ at $t$, what the Runge-Kutta-Fehlberg integrator finds is $\Omega$ at the next instant $t'$.
Updating $n$, $r$ and $\omega$ using $\Omega$ at $t'$ immediately follows. Updating the climate and ocean status at $t'$ starts in the next loop.
The above procedure is repeated for every instant until the final time.

%
%

\section{Results}\label{sec-results}

\subsection{Pre-analysis of tidal evolution}\label{sec-analysis}

\begin{figure}
\centering
\includegraphics[width=8cm]{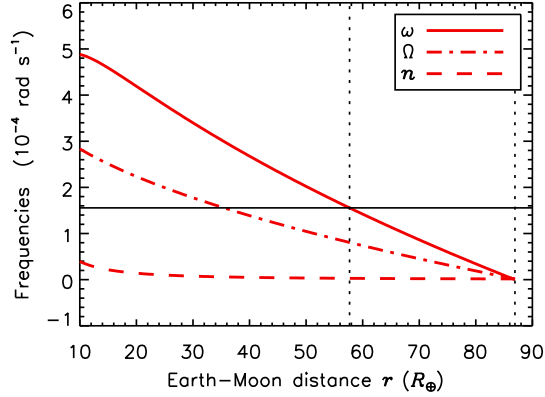}
\caption{Variations of tidal frequency $\omega$ (solid red curve), Earth's rotational speed $\Omega$ (dash-dotted red curve) and lunar orbital speed $n$ (dashed red curve) as functions of Earth-Moon distance $r$. The mean oceanic natural frequency $\bar{\sigma} = 1.555 \times 10^{-4}$\,rad~s$^{-1}$ (horizontal solid line) is illustrated for reference. The resonance distance $r_{\rm{res}} = 57.7\,R_{\oplus}$ where $\omega = \sigma$ and the synchronous distance $r_{\rm{syn}} = 86.9\,R_{\oplus}$ where $\Omega = n$ (left and right vertical dotted lines) are indicated.}
\label{fig-analysis}
\end{figure}

We first present a pre-analysis of the general trend of tidal evolution based on our model, for that helps in explaining the initialization of the numerical simulations.
According to our tidal evolution model, Earth's rotational speed $\Omega$ and the lunar orbital speed $n$ can be expressed as functions of the Earth-Moon distance $r$, and so can the tidal frequency $\omega$ (Sect.~\ref{sec-tidal}). If the oceanic natural frequency $\sigma$ is constant, the resonance distance $r_{\rm{res}}$, where the tidal resonance occurs as $\omega \approx \sigma$ (for a nonzero dissipation), can be then predicted without simulations.

Figure~\ref{fig-analysis} shows that as $r$ increases from $10\,R_{\oplus}$, both $\Omega$ and $n$ decrease, which means both one day $2\pi/\Omega$ and one month $2\pi/n$ lengthens. Their difference diminishes until $\Omega = n$, where Earth begins to synchronously rotate (one day is as long as one month) just like the Moon has been doing in reality (though Moon's rotation is neglected in our model) and the tidal evolution ends. According to the constants we set, that happens at $r_{\rm{syn}} = 86.9\,R_{\oplus}$.
In addition, as $r$ increases, $\omega$ also keeps decreasing until $\omega = 0$, when the resulting tidal torque and dissipation become zero. If the oceanic frequency is fixed at $\bar{\sigma} = 1.555 \times 10^{-4}$\,rad~s$^{-1}$, the tidal resonance is found to occur at $r_{\rm{res}} \approx 57.7\,R_{\oplus}$, slightly smaller than the present Earth-Moon distance $r_{\rm{p}} = 60.3\,R_{\oplus}$. This prediction is in accordance with the reality that the oceanic response has been near $M_2$ resonance currently.

Near $r_{\rm{res}}$, the tidal evolution should greatly speed up with dissipation of the total energy enhanced largely. The rapid decrease in $\omega$ at that time gives rise to the rapid pass through resonance. However, if $\sigma$ is varying, conditions become complicated. Before $r_{\rm{res}}$, a decreasing $\sigma$ delays the resonance while an increasing $\sigma$ hastens it; after $r_{\rm{res}}$, the former extends it while the latter curtails it. Even multiple passes can arise.
Those conditions are beyond the scope of the present work. Therefore, although we focus on the near resonance condition, we will simulate periods a while before and after $r_{\rm{res}}$, instead of simulating the maximum period.

\subsection{Numerical simulations}\label{sec-simulations}

\begin{table}
\begin{center}
\caption[]{Initialization and Results of Numerical simulations.}\label{tab-simulations}
\begin{tabular}{lccccccccc}
    \hline\noalign{\smallskip}
    Case    & $\psi$    & $r_{\rm{i}}$      & $r_{\rm{f}}$      & $2\pi/\Omega_{\rm{i}}$    & $2\pi/\Omega_{\rm{f}}$    & $2\pi/n_{\rm{i}}$   & $2\pi/n_{\rm{f}}$     & $\omega_{\rm{i}}$         & $\omega_{\rm{f}}$    \\
            & ($\dg$)   & ($R_{\oplus}$)    & ($R_{\oplus}$)    & (h)                    & (h)                    & (d)     & (d)     & ($10^{-4}$\,rad~s$^{-1}$) & ($10^{-4}$\,rad~s$^{-1}$)    \\
    \hline\noalign{\smallskip}
    B0      &       & 57.43 & 57.61 & 21.44 & 21.59 & 25.39 & 25.51 & 1.571  & 1.560 \\
    B$+$    & $+90$ & \ldots& $+4 \times 10^{-6}$ & \ldots& $+3 \times 10^{-6}$ & \ldots& $+3 \times 10^{-6}$ & \ldots& $-2 \times 10^{-7}$ \\
    B$-$    & $-90$ & \ldots& $-6 \times 10^{-6}$ & \ldots& $-5 \times 10^{-6}$      & \ldots& $-4 \times 10^{-6}$ & \ldots& $+4 \times 10^{-7}$ \\
    A0      &       & 57.80 & 57.98 & 21.74 & 21.89 & 25.63 & 25.75 & 1.549 & 1.538 \\
    A$+$    & $+90$ & \ldots& $-12 \times 10^{-6}$& \ldots& $-9 \times 10^{-6}$ & \ldots& $-8 \times 10^{-6}$ & \ldots& $+7 \times 10^{-7}$\\
    A$-$    & $-90$ & \ldots& $-5 \times 10^{-6}$ & \ldots& $-4 \times 10^{-6}$ & \ldots& $-3 \times 10^{-6}$ & \ldots& $+2 \times 10^{-7}$\\
    \noalign{\smallskip}\hline
\end{tabular}
\end{center}
\tablecomments{\textwidth}{For Cases B0 and A0, $\psi$ is not needed, because $\varepsilon$ is fixed at $\bar{\varepsilon}$.
Initial values substituted by dots are the same as above.
Final values for Cases B$+$ and B$-$ (A$+$ and A$-$) are presented as the divergences from those for Case B0 (A0).}
\end{table}

\begin{figure}
\centering
\includegraphics[width=16cm]{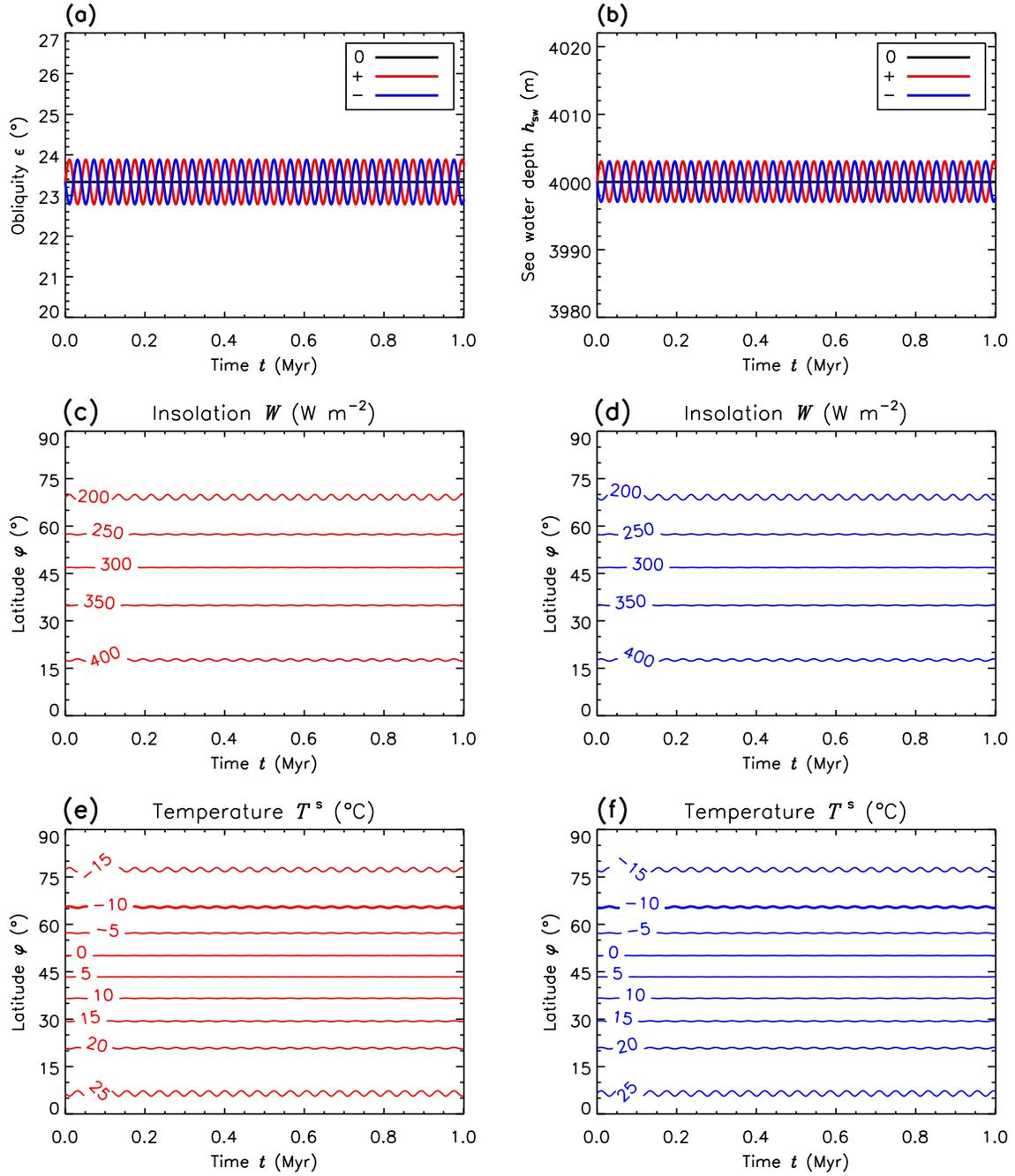}
\caption{Temporal variations of Earth's obliquity $\varepsilon$ (a), sea water depth $h_{\rm{sw}}$ (b), mean annual insolation function $W(\varphi)$ (c and d) and steady-state temperature field $T^{\rm{s}}(\varphi)$ (e and f) for both Cases B and A.
In panels a and b, the black lines, red curves and blue curves indicate Case 0, $+$ and $-$, respectively.
In panels c and e, contours are red to indicate Case $+$, while in d and f, those are blue to indicate Case $-$.}
\label{fig-climate}
\end{figure}

\begin{figure}
\centering
\includegraphics[width=16cm]{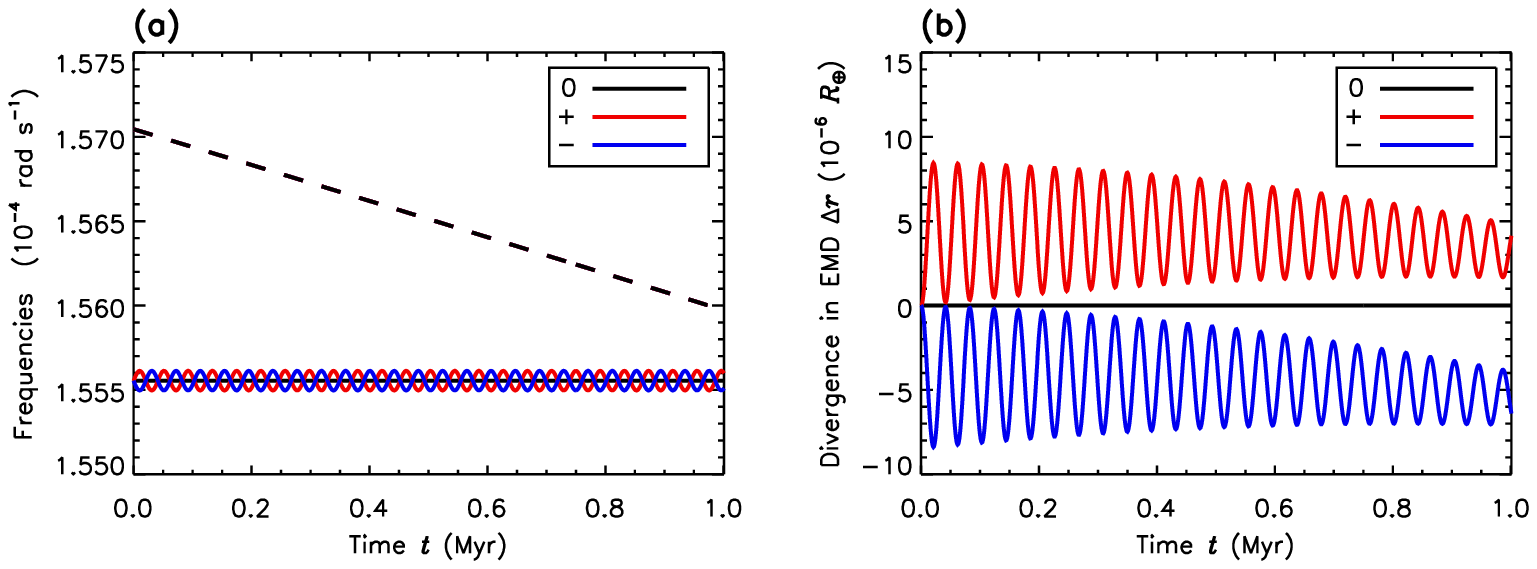}
\caption{Temporal variations of oceanic natural frequency $\sigma$, tidal forcing frequency $\omega$ (solid and dashed curves in panel a), and divergence in Earth-Moon distance $\Delta r$ (b) for Cases B.
In both panels, the colors black, red and blue indicate Cases B0, B$+$ and B$-$, respectively.
(In panel a, $\omega$ for Cases B$+$ and B$-$ are indistinguishable from B0)}
\label{fig-tidal-B}
\end{figure}

\begin{figure}
\centering
\includegraphics[width=16cm]{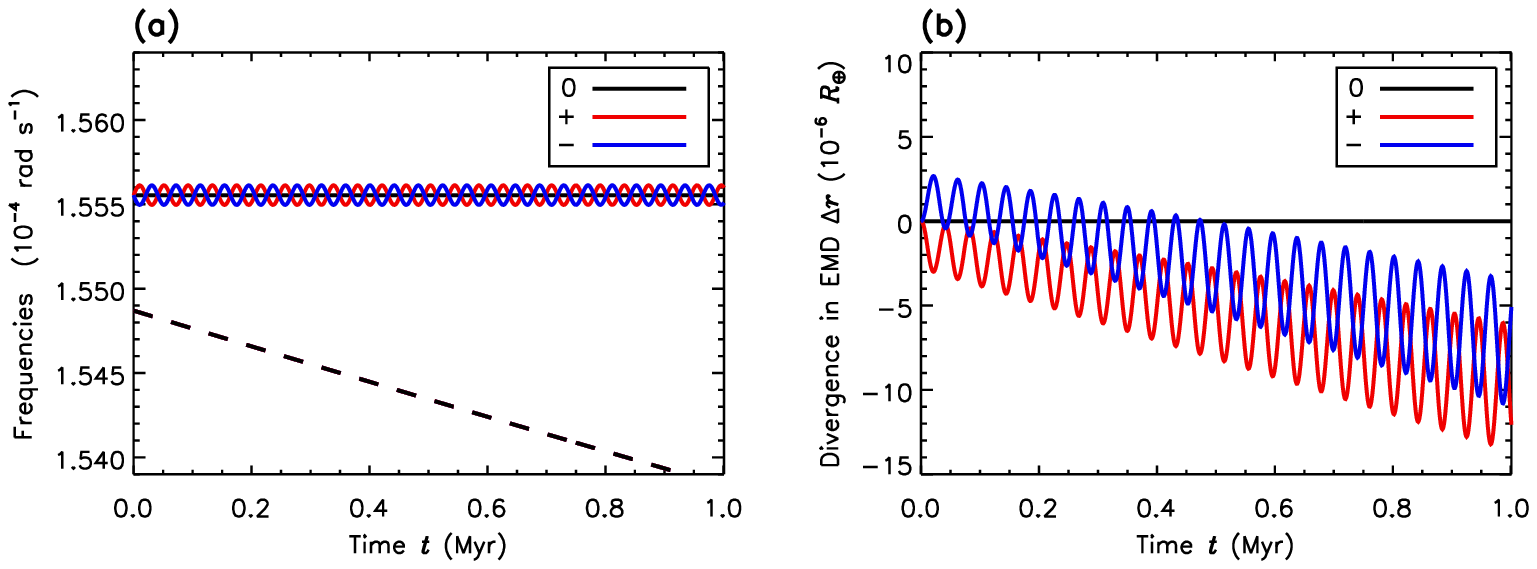}
\caption{Temporal variations of oceanic natural frequency $\sigma$, tidal forcing frequency $\omega$ (solid and dashed curves in panel a), and divergence in Earth-Moon distance $\Delta r$ (b) for Cases A.
In both panels, the colors black, red and blue indicate Cases A0, A$+$ and A$-$, respectively.
(In panel a, $\omega$ for Cases A$+$ and A$-$ are indistinguishable from A0)}
\label{fig-tidal-A}
\end{figure}

The initialization for our numerical simulations is shown in Table~\ref{tab-simulations}. Two sets of simulations, Cases B and Cases A ("before" and "after" resonance maximum respectively), are performed with initial Earth-Moon distance $r_{\rm{i}} = 57.43$ and 57.80\,$R_{\oplus}$, respectively, which are slightly smaller and larger than the resonance distance $r_{\rm{res}} \approx 57.7\,R_{\oplus}$ (Sect.~\ref{sec-analysis}). The initial values of the other two orbital parameters $\Omega_{\rm{i}}$ and $n_{\rm{i}}$ are derived from $r_{\rm{i}}$ (Sect.~\ref{sec-orb}). The corresponding initial tidal frequency $\omega_{\rm{i}} = 1.571 \times 10^{-4}$ and $1.549 \times 10^{-4}$\,rad~s$^{-1}$ for Cases B and A, respectively, just enclosing $\bar{\sigma} = 1.555 \times 10^{-4}$\,rad~s$^{-1}$, i.e., the oceanic frequency in the case of invariant $\varepsilon$.
Each set includes three cases: Case 0 acts as a control group with constant obliquity $\bar{\varepsilon} = 23.33\dg$; for Cases $+$ and $-$, $\varepsilon$ periodically varies with $\psi = +90\dg$ and $-90\dg$, respectively. The influence of the existence of climate change can thus be examined.
All the other parameters are set as described in Section~\ref{sec-method}.

We note that our aim in this work is to study the mechanism of climate influence but not the realistic history of lunar orbit. Therefore, although not comparable to the timescale of tidal evolution $10^9$\,yr, the simulation time $10^6$\,yr fits this aim, and it is not a problem that the cases with varying $\varepsilon$ are initialized at the same $r_{\rm{i}}$ as the case with constant $\varepsilon$. Furthermore, it is therefore reasonable to randomly set the phase of obliquity $\psi$ at any $r$. We choose $+90$ and $-90\dg$ in order to maximize the difference between Cases $+$ and $-$.

Cases B start and end before the resonance maximum (Table~\ref{tab-simulations}). As shown in Figure~\ref{fig-climate}a, $\varepsilon$ increases and decreases from $\bar{\varepsilon}$ at the beginning for Case B$+$ and Case B$-$, respectively, and periodically varies till the end. According to \cite{Loutre-2004}, the mean annual insolation $W$, being symmetrical between northern and southern hemispheres, varies in phase with $\varepsilon$ in the high latitudes but exactly out of phase in the low latitudes. These properties are completely exhibited in Figure~\ref{fig-climate}c and \ref{fig-climate}d. The steady-state temperature $T^{\rm{s}}$, as a response to insolation, has the same properties as $W$. As shown in Figure~\ref{fig-climate}e and \ref{fig-climate}f, located in the high latitudes, the iceline $\varphi_{\rm{f}}$ varies in phase with $\varepsilon$ for both Cases B$+$ and B$-$. Its mean $\bar{\varphi}_{\rm{f}} = 65.44\dg$ and the amplitude defined as the maximal deviation from the mean is $0.23\dg$.
Because the sea water depth $h_{\rm{sw}} \propto \sin\varphi_{\rm{f}}$ (Eq.~\ref{eq-h_sw}) and the oceanic frequency $\sigma \propto \sqrt{h_{\rm{sw}}}$ (Eq.~\ref{eq-omega_0}), they are in phase with $\varepsilon$ as well. For either Case B$+$ or B$-$, $h_{\rm{sw}}$ oscillates about $\bar{h}_{\rm{sw}}$ with an amplitude of 2.98\,m (Fig.~\ref{fig-climate}b), and $\sigma$ oscillates about $\bar{\sigma}$ with an amplitude of $0.001 \times 10^{-4}$\,rad~s$^{-1}$ (Fig.~\ref{fig-tidal-B}a).

The general trend of tidal evolution with constant $\sigma$ has been interpreted in Section~\ref{sec-analysis}. We now list the initial and final values of the lunar orbital parameters in Table~\ref{tab-simulations} and illustrate only the divergence of Earth-Moon distance $\Delta r$ for Cases B$+$ and B$-$ from Case B0 in Figure~\ref{fig-tidal-B}b in order to focus on the climate influence.
Three features are commonly observed for Cases B$+$ and B$-$.
First, $\Delta r$ varies in phase with $\sigma$ and thus in phase with $\varepsilon$. The reason is that during the pre-resonance time when $\omega > \sigma$, greater $\sigma$ means greater tidal dissipation, resulting in a leading evolution characterized by a larger $r$.
Second, the equilibrium point of the $\Delta r$ oscillation is not constant but seems to decrease at least on the near side of resonance maximum.
Third, the displacement of $\Delta r$ from the gradually decreasing equilibrium point expresses a positive correlation with the difference between $\omega$ and $\sigma$.
One distinction between Cases B$+$ and B$-$ is that the mean of $\Delta r$ for the former is larger than the latter. We attribute this distinction to $\psi$ and the beginning behavior of $\Delta r$ it determines: the beginning increase/decrease in $\Delta r$ for Case B$+$/B$-$ contributes to a lead/drop lasting for the whole simulation time.

Cases A start later than the resonance maximum (Table~\ref{tab-simulations}). The evolutions of climate and ocean state ($\varepsilon$, $W$, $T^{\rm{s}}$, $\varphi_{\rm{f}}$, $h_{\rm{sw}}$ and $\sigma$) for Cases A are totally the same as those for Cases B.
However, because $\omega < \sigma$ during the post-resonance time (Fig.~\ref{fig-tidal-A}a), contrary to Cases B, greater $\sigma$ means smaller dissipation and thus a smaller $r$. As shown in Figure~\ref{fig-tidal-A}b, $\Delta r$ for either Case A$+$ or A$-$ is therefore exactly out of phase with $\sigma$.
The second and third features for Cases B, i.e., the general trend of decreasing and positive correlation between $\Delta r$ displacement and the difference between $\omega$ and $\sigma$, also match Cases A.
In addition, $\psi$ again results in a larger mean of $\Delta r$ for the case where $\Delta r$ increases at the beginning (A$-$) than where it decreases (A$+$). It is worth mentioning that both as the case with the larger mean $\Delta r$, B$+$ holds a positive $\Delta r$ for the whole time, whereas A$-$ holds for only about $10^5$\,yr. Considering the general decreasing trend, the fact that B$+$ holds a lead over B0 is probably a temporary effect.

In summary, the features of our climate-influenced tidal evolution (characterized by $\Delta r$ whose positive value means a lead over the evolution with climate unchanged and negative value means a lag) are
\begin{enumerate}
  \item Given that iceline $\varphi_{\rm{f}}$ is in high latitudes (so that $\sigma$ is in phase with $\varepsilon$), $\Delta r$ varies in phase with $\varepsilon$ during the pre-resonance time but exactly out of phase during the post-resonance time.
  \item Despite oscillation, the general trend of $\Delta r$ is decreasing.
  \item The displacement of $\Delta r$ oscillation is in positive correlation with the difference between $\omega$ and $\sigma$.
  \item As a whole, $\Delta r$ oscillation is shifted upwards or downwards as $\Delta r$ increases or decreases at the beginning.
\end{enumerate}

\section{Discussion}\label{sec-discussion}

Based on our conceptual coupled model of climate and tidal evolution (Sect.~\ref{sec-model}), we carried out numerical simulations of the near-resonance tidal evolution for an equatorial circular lunar orbit with Earth's obliquity $\varepsilon$ periodically varying (Sect.~\ref{sec-simulations}).
Thus, the climate influence on the tidal evolution via ocean is verified. Our conclusions in terms of the influence mechanism are qualitative. The main conclusion is that compared to the case that the climate is invariant, varying climate slows down the evolution accompanied by oscillations. Furthermore, the oscillation is in phase and exactly out of phase with $\varepsilon$ before and after the resonance maximum, respectively; and can be enlarged as the difference between the oceanic frequency $\sigma$ and the tidal frequency $\omega$ increases.

The above conclusions should be applied with caution. This is not only because of the idealization and the existence of multiple parameters of the model, but also because the simulations are only done for a short instant near the resonance maximum of the whole lunar tidal evolution.

Though we focus on the mechanism of climate influence in this work, it should be pointed out that the absolute differences in final orbital parameters found between the cases with varying and invariant climates (Table~\ref{tab-simulations}) are insignificant indeed. However, it is still hasty to conclude that the influence of climate change can be neglected.
On one hand, the timescale studied here, $10^6$\,yr, is very short compared with the timescale of tidal evolution, $10^9$\,yr. If the evolution keeps slowing down with climate varying, the secular accumulation may make a difference.
On the other hand, the variations of the climate and ocean state produced here are not as big as in reality. The maximal drop of the sea water depth in simulations is 6\,m, whereas the sea level drop during the last glacial maximum relative to the present is about 130\,m (\citealt{Clark-2009}). If a more realistic model is used, the influence will be enhanced.

One important effect that can enhance the climate influence is the "ice-albedo feedback." In the current model, though the ice sheet on the continent is considered, the coalbedo $\tilde{\alpha}$ has no dependence on iceline $\varphi_{\rm{f}}$. A more realistic way, for example, is to multiply $\tilde{\alpha}$ by $1/2$ in latitudes higher than $\varphi_{\rm{f}}$ (\citealt{Mengel-1988}). In this case, as the ice cover spreads, the planetary coalbedo and thus the absorbed solar radiation diminishes, leading to a further drop in temperature accompanied by the spread of ice cover. In other words, a slight change in solar radiation can cause an abrupt climate transition (\citealt{North-1984}). This nonlinear feedback mechanism will be introduced in our future model.

%
%

Potential subjects of future works include improving the model, determining the quantitative correlation between climate variation and the rate of tidal evolution, and generalizing the model to other planet-satellite systems.
In addition, considering that it is the normal modes of the liquid part of the Earth that can largely be excited when tidal resonance occurs, the tidal evolutions of terrestrial planets perturbed by companions in exosolar systems (e.g., \citealt{Dong-2013}) may also need further investigations when oceans or liquid cores (\citealt{Liu-2018}) are present.

\begin{acknowledgements}
We thank the referee for constructive comments. We are grateful that Professor Kwang-Yul Kim helped in climate simulations. The code of tidal evolution was developed in Nanjing University when the author Nan Wang was under the supervision of Professor Ji-Lin Zhou, and further improved in Zhejiang University.
This work was funded by the Natural Science Foundation of Zhejiang Province (LR16E090001), the National Key Research and Development Program of China (Grant No. 2017YFC0305905), and NSFC-Zhejiang Joint Fund for the Integration of Industrialization and Informatization (Grant No. U1709204).
\end{acknowledgements}

\bibliographystyle{raa}
\bibliography{RAA-2018-0288}

\begin{thebibliography}{41}
\providecommand\natexlab[1]{#1}
\providecommand\JournalTitle[1]{#1}

\bibitem[Berger(1988)]{Berger-1988}
Berger, A. 1988, Reviews of Geophysics, 26, 624

\bibitem[Berger(2012)]{Berger-2012}
Berger, A. 2012, A Brief History of the Astronomical Theories of Paleoclimates,
  ed. A.~Berger, F.~Mesinger, \& D.~Sijacki, Climate Change: Inferences from
  Paleoclimate and Regional Aspects, ed. A.~Berger, F.~Mesinger, \& D.~Sijacki
  (Vienna: Springer Vienna), 107

\bibitem[Berger(1978)]{Berger-1978-JAS}
Berger, A.~L. 1978, Journal of the Atmospheric Sciences, 35, 2362

\bibitem[Berger \& Loutre(1991)]{Berger-1991}
Berger, A., \& Loutre, M.~F. 1991, Quaternary Science Reviews, 10, 297

\bibitem[Berger \& Loutre(1992)]{Berger-1992}
Berger, A., \& Loutre, M.~F. 1992, Earth and Planetary Science Letters, 111,
  369

\bibitem[Berger {et~al.}(2010)]{Berger-2010}
Berger, A., Loutre, M.-F., \& Yin, Q. 2010, Quaternary Science Reviews, 29,
  1968

\bibitem[Bills \& Ray(1999)]{Bills-1999}
Bills, B.~G., \& Ray, R.~D. 1999, Geophysical Research Letters, 26, 3045

\bibitem[Budyko(1969)]{Budyko-1969}
Budyko, M.~I. 1969, Tellus, 21, 611

\bibitem[Clark {et~al.}(2009)]{Clark-2009}
Clark, P.~U., Dyke, A.~S., Shakun, J.~D., {et~al.} 2009, Science, 325, 710

\bibitem[Dong \& Ji(2013)]{Dong-2013}
Dong, Y., \& Ji, J. 2013, Monthly Notices of the Royal Astronomical Society,
  430, 951

\bibitem[Egbert {et~al.}(2004)]{Egbert-2004}
Egbert, G.~D., Ray, R.~D., \& Bills, B.~G. 2004, Journal of Geophysical
  Research: Oceans, 109, C03003

\bibitem[Goldreich(1966)]{Goldreich-1966}
Goldreich, P. 1966, Reviews of Geophysics, 4, 411

\bibitem[Green {et~al.}(2009)]{Green-2009}
Green, J. A.~M., Green, C.~L., Bigg, G.~R., {et~al.} 2009, Geophysical Research
  Letters, 36, 5

\bibitem[Griffiths \& Peltier(2009)]{Griffiths-2009}
Griffiths, S.~D., \& Peltier, W.~R. 2009, Journal of Climate, 22, 2905

\bibitem[Halliday(2008)]{Halliday-2008}
Halliday, A.~N. 2008, Philosophical Transactions of the Royal Society A:
  Mathematical, Physical and Engineering Sciences, 366, 4163

\bibitem[Hansen(1982)]{Hansen-1982}
Hansen, K.~S. 1982, Reviews of Geophysics, 20, 457

\bibitem[Haynes(2017)]{Haynes-2017}
Haynes, W.~M., ed. 2017, CRC Handbook of Chemistry and Physics, 97th edn. (Boca
  Raton, FL: CRC Press)

\bibitem[Huybers \& Tziperman(2008)]{Huybers-2008}
Huybers, P., \& Tziperman, E. 2008, Paleoceanography, 23, PA1208

\bibitem[Kagan \& Maslova(1994)]{Kagan-1994}
Kagan, B.~A., \& Maslova, N.~B. 1994, Earth, Moon, and Planets, 66, 173

\bibitem[Lambeck {et~al.}(2014)]{Lambeck-2014}
Lambeck, K., Rouby, H., Purcell, A., Sun, Y., \& Sambridge, M. 2014,
  Proceedings of the National Academy of Sciences, 111, 15296

\bibitem[Laskar(1988)]{Laskar-1988}
Laskar, J. 1988, Astronomy and Astrophysics, 198, 341

\bibitem[Lin \& North(1990)]{Lin-1990}
Lin, R.~Q., \& North, G.~R. 1990, Climate Dynamics, 4, 253

\bibitem[Liu \& Li(2018)]{Liu-2018}
Liu, M., \& Li, L.-G. 2018, Research in Astronomy and Astrophysics, 18, 23

\bibitem[Loutre {et~al.}(2004)]{Loutre-2004}
Loutre, M.-F., Paillard, D., Vimeux, F., \& Cortijo, E. 2004, Earth and
  Planetary Science Letters, 221, 1

\bibitem[McGehee \& Lehman(2012)]{McGehee-2012}
McGehee, R., \& Lehman, C. 2012, SIAM Journal on Applied Dynamical Systems, 11,
  684

\bibitem[Mengel {et~al.}(1988)]{Mengel-1988}
Mengel, J.~G., Short, D.~A., \& North, G.~R. 1988, Climate Dynamics, 2, 127

\bibitem[Munk(1968)]{Munk-1968}
Munk, W. 1968, Quarterly Journal of the Royal Astronomical Society, 9, 352

\bibitem[Munk(1997)]{Munk-1997}
Munk, W. 1997, Progress in Oceanography, 40, 7

\bibitem[Munk \& Wunsch(1998)]{Munk-1998}
Munk, W., \& Wunsch, C. 1998, Deep Sea Research Part I: Oceanographic Research
  Papers, 45, 1977

\bibitem[Murray \& Dermott(1999)]{Murray-1999}
Murray, C.~D., \& Dermott, S.~F. 1999, Solar system dynamics (Cambridge
  university press)

\bibitem[North(1984)]{North-1984}
North, G.~R. 1984, Journal of the Atmospheric Sciences, 41, 3390

\bibitem[North \& Jr.(1979)]{North-1979}
North, G.~R., \& Jr., J. A.~C. 1979, Journal of the Atmospheric Sciences, 36,
  1189

\bibitem[North \& Kim(2017)]{North-2017}
North, G.~R., \& Kim, K.-Y. 2017, Energy Balance Climate Models, Wiley Series
  in Atmospheric Physics and Remote Sensing (Wiley-VCH Verlag GmbH \& Co. KGaA)

\bibitem[Sellers(1969)]{Sellers-1969}
Sellers, W.~D. 1969, Journal of Applied Meteorology, 8, 392

\bibitem[Thomas \& S{\"u}ndermann(1999)]{Thomas-1999}
Thomas, M., \& S{\"u}ndermann, J. 1999, Journal of Geophysical Research:
  Oceans, 104, 3159

\bibitem[Touma \& Wisdom(1994)]{Touma-1994}
Touma, J., \& Wisdom, J. 1994, The Astronomical Journal, 108, 1943

\bibitem[Wagner \& Eisenman(2015)]{Wagner-2015}
Wagner, T. J.~W., \& Eisenman, I. 2015, Journal of Climate, 28, 3998

\bibitem[Webb(1980)]{Webb-1980}
Webb, D.~J. 1980, Geophysical Journal International, 61, 573

\bibitem[Webb(1982{\natexlab{a}})]{Webb-1982-GJI-68}
Webb, D.~J. 1982{\natexlab{a}}, Geophysical Journal International, 68, 689

\bibitem[Webb(1982{\natexlab{b}})]{Webb-1982-GJI-70}
Webb, D.~J. 1982{\natexlab{b}}, Geophysical Journal International, 70, 261

\bibitem[Yokoyama {et~al.}(2000)]{Yokoyama-2000}
Yokoyama, Y., Lambeck, K., De~Deckker, P., Johnston, P., \& Fifield, L.~K.
  2000, Nature, 406, 713

\end{thebibliography}

\label{lastpage}
\end{document}